\documentclass[twoside]{article}
\usepackage{fleqn}
\usepackage{espcrc2}
\usepackage{epsf}
\usepackage{graphics}
\newcommand{\be} {\begin{equation}}
\newcommand{\ee} {\end{equation}}
\newcommand{\bea}{\begin{eqnarray}}
\newcommand{\eea}{\end{eqnarray}}
\newcommand{\bi}{\bibitem}
\newcommand{\dst}[1]{\bar{s} #1 d \, \bar{s} #1 d}

\def\lvec#1{\setbox0=\hbox{$#1$}
    \setbox1=\hbox{$\scriptstyle\leftarrow$}
    #1\kern-\wd0\smash{
    \raise\ht0\hbox{$\raise1pt\hbox{$\scriptstyle\leftarrow$}$}}
    \kern-\wd1\kern\wd0}
\def\rvec#1{\setbox0=\hbox{$#1$}
    \setbox1=\hbox{$\scriptstyle\rightarrow$}
    #1\kern-\wd0\smash{
    \raise\ht0\hbox{$\raise1pt\hbox{$\scriptstyle\rightarrow$}$}}
    \kern-\wd1\kern\wd0}

\def\cancel#1#2{\ooalign{$\hfil#1\mkern1mu/\hfil$\crcr$#1#2$}}
\def\slash#1{\mathpalette\cancel{#1}}

\newcommand{\inc}[1]{\resizebox{7.5cm}{!}{\rotatebox{-90}{\includegraphics{#1}}}}

\title{\vspace{-4.0cm} 
\begin{flushright}
{\normalsize\tt HET preprint}\\
\vspace{-0.3cm}
{\normalsize\tt BNL-HET-99/xx}\\
\end{flushright}
\vspace*{3.0cm}
Non-Perturbative Renormalisation using Domain Wall Fermions
}
\author{C.~Dawson \address{Physics Department, Brookhaven
National Laboratory, PO Box 5000, Upton, NY 11973-5000, USA\\
cdawson@bnl.gov}
\thanks{This manuscript has been authored under
contract number DE-AC02-98CH10886 with the U.S.~Department of Energy.
Work done in collaboration with T. Blum, P. Chen,
N. Christ, M. Creutz, C. Dawson, G. Fleming, A. Kaehler, T. Klassen,
C. Malureanu, R. Mawhinney, S. Ohta, S. Sasaki, G. Siegert, C. Sui,
A. Soni, M. Wingate, P. Vranas, L.  Wu, and Yu. Zhestkov .} }
\begin{document}
\pagestyle{empty}  


%
%
\begin{abstract}  
The viability of the Non-Perturbative Renormalisation (NPR) method of
the Rome/Southampton group is studied, for the first time, in the
context of domain wall fermions. The procedure is used to extract the
renormalisation coefficients of the various quark bilinears, as well
as the four-fermion operators relevant to the $\Delta S=2$ effective
Hamiltonian. The renormalisation of the $\Delta S=1$ Hamiltonian is
also discussed.
\end{abstract}
\maketitle
\section{INTRODUCTION}
Domain Wall Fermions \cite{ref:shamir}(DWF) are an attractive approach
to Lattice QCD that allows exact vector and axial symmetry at finite
lattice spacing, the latter in the limit of an infinite fifth
dimension. This property greatly simplifies the renormalisation
properties of lattice operators. However, realistic simulations are
performed with a fifth dimension of finite extent (16 sites for the results
presented below) and so potential breaking of chiral symmetry must be
considered \cite{ref:st}.

Below a brief overview of a preliminary study of the 
renormalisation properties of several operators using the
regularisation independent technique of
the Rome/Southampton group (RI/MOM) \cite{ref:gen_npr} is given. 
\section{QUARK PROPAGATOR}
\begin{figure}
\inc{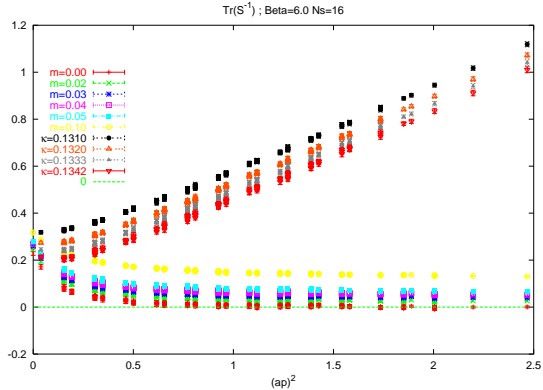}
\vspace{-0.8cm}
\caption{$Z_q M_{RI}$}
\vspace{-0.8cm}
\label{fig:mass}
\end{figure}
The RI/MOM scheme requires the calculation of the operators of
interest between external quark states, in a fixed gauge, at high
momenta. For many quantities of interest this requires only knowlege
of the momentum space quark propagator from a fixed origin, $S(p)$, and so
simulations are relatively ``inexpensive''.  

From just the propagator itself the quark renormalisation factor, $Z_q^{1/2}$,
and, more interestingly, the quark mass may be calculated.
%
An RI/MOM scheme mass may be defined from the inverse propagator as
\be
M_{RI} 
= 
\frac{1}{Z_q} \, \frac{Tr S^{-1}(p^2)}{12} 
= 
Z_m \, m \ .
\label{eq:zm}
\ee 
This definition of the mass is extremely sensitive to the presence of explicit
breaking of chiral symmetry at $O(a)$, in which case an additive
renormalisation is needed \cite{ref:sst,ref:vit}. Fig. \ref{fig:mass}
contrasts $Z_q M_{RI}$ for both non-perturbatively improved Wilson
fermions~\footnote{Thanks to A.P.E. for this data} and DWF, showing
explicitly the good chiral symmetry of the DWF,with no additive term
apparent within errors.
Both lattices had dimensions $16^3 \times 32$ with $\beta=6.0$ and the
results shown are for $37$ and $52$ gauge configurations respectively.
%
%
\section{FERMION BILINEARS}
The renormalisation factors for the flavour non-singlet fermion bilinears in
the RI scheme may be calculated from~\cite{ref:bils}
\be
\frac{Z_\Gamma}{ 12 \, Z_q} Tr \left[ \Gamma \langle  q | \bar{\psi} \Gamma \psi
\,
| \bar{q} \rangle_{AMP} \right] |_{p^2=\mu^2} = 1 \ ,
\label{eq:bilinear}
\ee
with $\Gamma = \left\{ S, V, P, A, T \right\}$ .

\begin{figure}
\vspace{-0.5cm}
\inc{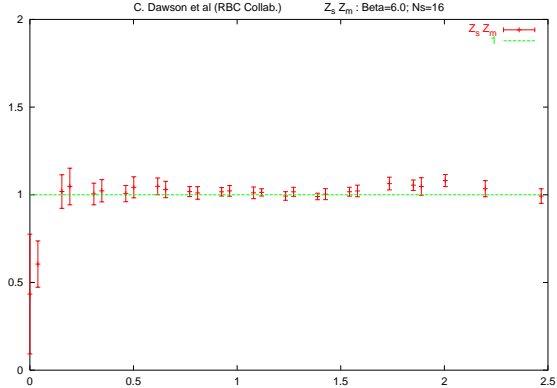}
\vspace{-0.8cm}
\caption{$Z_S \times Z_M$}
\vspace{-0.5cm}
\label{fig:zs_times_zm}
\end{figure}

Values for these quantities and comparison to perturbation theory will appear
in a forthcoming paper. For the moment only properties relevant for the
extraction of the quark masses will be mentioned. In particular, if both the
vector and axial symmetries are respected, the relation $Z_S=1/Z_m$, 
should hold. Fig. \ref{fig:zs_times_zm} shows $Z_s \times Z_m$ vs $(ap)^2$. The
fact that this relation holds so well both confirms the
predictions of perturbation theory \cite{ref:aoki} and gives confidence in the
method used to extract the strange quark mass in \cite{ref:matt}, where more
details may be found. 
\section{$\Delta S=2$ HAMILTONIAN}
\begin{figure}
\vspace{-0.5cm}
\inc{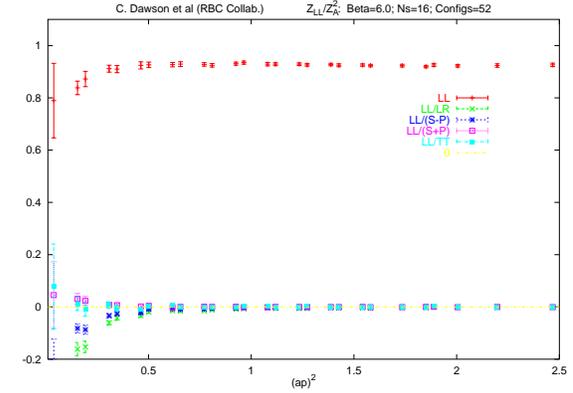}
\vspace{-0.8cm}
\caption{$Z_{\Delta S =2} / Z_A^2$}
\vspace{-0.5cm}
\label{fig:zll}
\end{figure}
The parity conserving part of the $\Delta S=2$ Hamiltonian is proportional
to
\bea
O_{VV+AA}
=
\dst{\gamma_\mu \gamma_5}
+
\dst{\gamma_\mu} \, .
\eea
In the continuum this renormalises multiplicatively, but in
the presence of explicit chiral symmetry breaking four other operators may mix;
$O_{VV-AA}$, $O_{SS-PP}$, $O_{SS+PP}$ and  $O_{TT}$ \cite{ref:fferm}.

An RI-scheme renormalisation condition may be imposed on this set of operators
by constructing the matrix elements of $O_i$ between external quark states,
\begin{equation}
\Lambda^{i}_{\alpha \beta \gamma \delta}
=
\langle 
        \, 0 \, | 
        \ O_i \  
        \psi_\alpha \bar{\psi}_\beta \psi_\gamma \bar{\psi}_\delta  
        | \, 0 \, \rangle_{amp} \, ,
\label{eq:lambda}
\end{equation}
taking the projection of these matrix elements with a complete
parity conserving basis in spinor space,
\be
P_{\Gamma\otimes \Gamma} \left[ \Lambda \right]
= 
\Gamma^i_{\alpha \beta}
\Gamma^i_{\gamma \delta}
\
\Lambda^j_{\alpha \beta \gamma \delta} 
\label{eq:proj}
\ee
and then requiring that, for renormalised operators, this be equal to its free case value,
\be
Z_{i\,j}\, 
P_{\left(\Gamma \otimes \Gamma\right)_{j}} 
\left[ \Lambda^{k} \right] = F_{i\,k} \ .
\ee
Fig. \ref{fig:zll} shows the overall renormalisation and mixings for
$O_{VV+AA}$, and as can be seen, in a simulation with a fifth dimension of 16
points there is essentially no mixing from chirally disallowed operators. This
may be contrasted to the case for Wilson fermions in which mixings are of
order $10 \%$ \cite{ref:fferm}.  It should be noted, that even with a
fifth dimension of only four sites, while the mixing with chirally disallowed
operators becomes appreciable, the magnitude of the mixing coefficients is
smaller than that for Wilson fermions.
\section{$\Delta S=1$ HAMILTONIAN}
\begin{figure}
\vspace{-0.5cm}
\begin{center}
\resizebox{2cm}{!}{\includegraphics{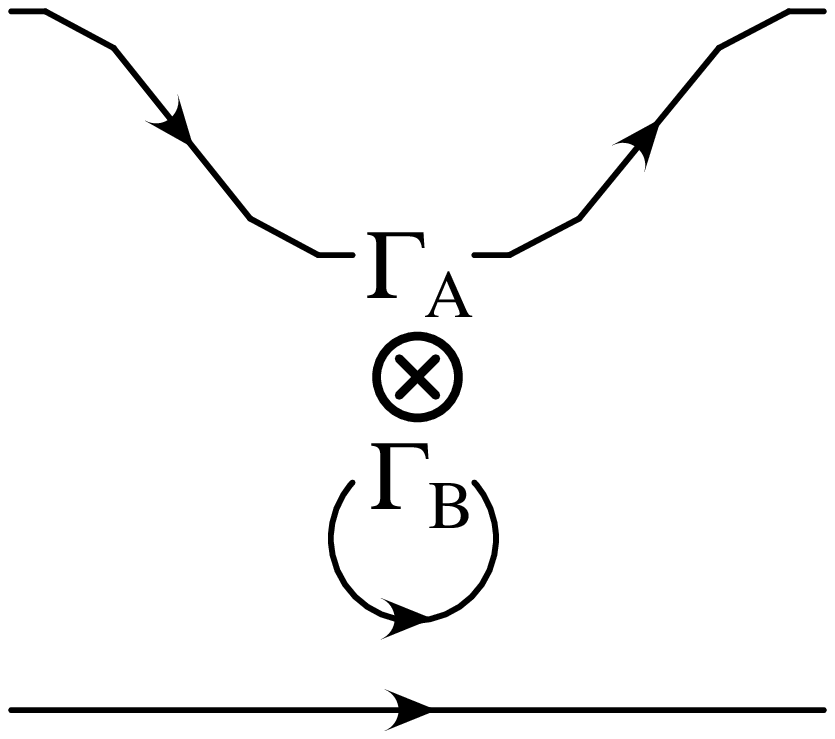}}
\vspace{-0.8cm}
\caption{``eye'' diagram}
\vspace{-1.0cm}
\label{fig:eye}
\end{center}
\end{figure}

The renormalisation of the parity conserving part of the $\Delta S=1$
Hamiltonian is a much more challenging problem on the lattice than that of the
fermion bilinears or the $\Delta S=2$ Hamiltonian, but also of much greater
interest phenomonoligically \cite{ref:soni}. The difficulty in evaluating the
renormalisation conditions for this Hamiltonian can be attributed to the
appearance of ``eye'' diagrams such as that shown in Fig. \ref{fig:eye}.
These diagram introduce two problems. 

Firstly they cannot be constructed from
merely the momentum space quark propagator from a single origin, as they
require a completely disconnected quark propagator transformed into momentum
space, \be S(p,q) = \sum_{x,y} e^{-ip.x} S(x,y) e^{iq.y} \, .  \ee This may be
calculated by inverting the Dirac equation using a fixed momentum source, but
this ``costs'' a matrix inversion for every momenta, q, needed. For this
reason only four fixed momenta have been used during this exploratory study.

Secondly such diagrams allow mixing with lower dimensional operators,
including, but not limited to, off-shell ``Equation of Motion'' operators
\cite{ref:me}, such as; \be \bar{s}(-\lvec{\slash{D}}+m \, )d + \bar{s}
(\rvec{\slash{D}}+m ) d \, .  \ee All such operators must be subtracted (or
shown to be irrelevant) before the equivalent of
Eq.(\ref{eq:proj})~\footnote{A complete basis of projectors in both spin and
flavour space must be used in this case.} may be applied. This may be done in
principle by, for example, studying the matrix elements of the effective
Hamiltonian between external $s$ and $d$ quark states \cite{ref:mm}, but in
practice this procedure only seems possible in the case of a lattice action
that respects chiral symmetry due to the proliferation of possible operator
mixings otherwise.

\begin{figure}
\vspace{-0.5cm}
\inc{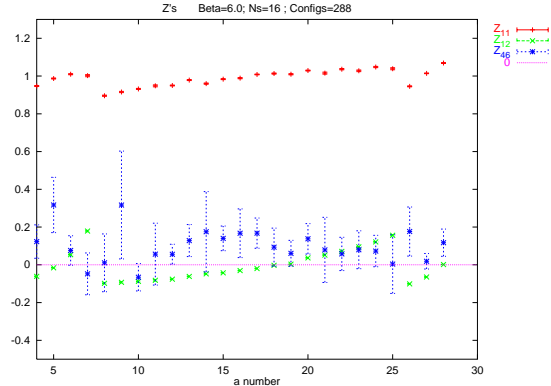}
\vspace{-0.8cm}
\label{fig:zds1}
\caption{Non-subtracted renormalisation factors}
\vspace{-0.5cm}
\end{figure}

Fig. \ref{fig:zds1} shows three sample renormalisation factors, on a $16^3
\times 32 \times 16$ lattice, using $288$ gauge configurations, for
$m=0.04$. The subtraction of the lower dimensional operators has
currently not been completed, but preliminary results suggest that the
effects of these terms are small. 

\section{CONCLUSIONS}
A preliminary study of the RI/MOM scheme applied to DWF has performed. The
results seem encouraging, especially in the (low) level of chiral symmtery
breaking exibited.

\end{document}